\documentclass[10pt, conference, compsocconf]{IEEEtran}

\usepackage{amsmath}
\usepackage{amssymb}
\usepackage{graphicx}

\catcode`@=11
\chardef\mathlig@atcode\count255

\def\actively#1#2{\begingroup\uccode`\~=`#2\relax\uppercase{\endgroup#1~}}
\def\mathlig@gobble{\afterassignment\mathlig@next@cmd\let\mathlig@next= }
\def\mathlig@delim{\mathlig@delim}
\def\mathlig@defcs#1{\expandafter\def\csname#1\endcsname}
\def\mathlig@let@cs#1#2{\expandafter\let\expandafter#1\csname#2\endcsname}
\def\mathlig@appendcs#1#2{\expandafter\edef\csname#1\endcsname{\csname#1\endcsname#2}}

\def\mathlig#1#2{\mathlig@checklig#1\mathlig@end\mathlig@defcs{mathlig@back@#1}{#2}\ignorespaces}

\def\mathlig@checklig#1#2\mathlig@end{%
 \expandafter\ifx\csname mathlig@forw@#1\endcsname\relax
 \expandafter\mathchardef\csname mathlig@back@#1\endcsname=\mathcode`#1%
 \mathcode`#1"8000\actively\def#1{\csname mathlig@look@#1\endcsname}%
 \mathlig@dolig#1\mathlig@delim
\fi
\mathlig@checksuffix#1#2\mathlig@end
}

\def\mathlig@checksuffix#1#2\mathlig@end{%
\ifx\mathlig@delim#2\mathlig@delim\relax\else\mathlig@checksuffix@{#1}#2\mathlig@end\fi
}
\def\mathlig@checksuffix@#1#2#3\mathlig@end{%
\expandafter\ifx\csname mathlig@forw@#1#2\endcsname\relax\mathlig@dosuffix{#1}{#2}\fi
\mathlig@checksuffix{#1#2}#3\mathlig@end
}

\def\mathlig@dosuffix#1#2{%
\mathlig@appendcs{mathlig@toks@#1}{#2}%
\mathlig@dolig{#1}{#2}\mathlig@delim
}

\def\mathlig@dolig#1#2\mathlig@delim{%
 \mathlig@defcs{mathlig@look@#1#2}{%
 \mathlig@let@cs\mathlig@next{mathlig@forw@#1#2}\futurelet\mathlig@next@tok\mathlig@next}%
 \mathlig@defcs{mathlig@forw@#1#2}{%
  \mathlig@let@cs\mathlig@next{mathlig@back@#1#2}%
  \mathlig@let@cs\checker{mathlig@chck@#1#2}%
  \mathlig@let@cs\mathligtoks{mathlig@toks@#1#2}%
  \expandafter\ifx\expandafter\mathlig@delim\mathligtoks\mathlig@delim\relax\else
  \expandafter\checker\mathligtoks\mathlig@delim\fi
  \mathlig@next
 }%
 \mathlig@defcs{mathlig@toks@#1#2}{}%
 \mathlig@defcs{mathlig@chck@#1#2}##1##2\mathlig@delim{%
  \ifx\mathlig@next@tok##1%
   \mathlig@let@cs\mathlig@next@cmd{mathlig@look@#1#2##1}\let\mathlig@next\mathlig@gobble
  \fi 
  \ifx\mathlig@delim##2\mathlig@delim\relax\else
   \csname mathlig@chck@#1#2\endcsname##2\mathlig@delim
  \fi
 }%
 \ifx\mathlig@delim#2\mathlig@delim\else
  \mathlig@defcs{mathlig@back@#1#2}{\csname mathlig@back@#1\endcsname #2}%
 \fi
}%

\mathchardef\mathlig@paren\mathcode`(

\begingroup \catcode`\"=12
\gdef\resetMathstrut@{%
  \setbox\z@\hbox{%
    \mathchardef\@tempa\mathlig@paren\relax
    \def\@tempb##1"##2##3{\the\textfont"##3\char"}%
    \expandafter\@tempb\meaning\@tempa \relax
  }%
  \ht\Mathstrutbox@\ht\z@ \dp\Mathstrutbox@\dp\z@
}
\endgroup

\catcode`@\mathlig@atcode

\mathchardef\ordinarypar\mathcode`\|

\mathchardef\ordinarycolon\mathcode`\:

\newcommand{\pinull}{\mathbf{0}}
\newcommand{\opm}[2]{\overline{#1} \, #2 }
\newcommand{\ipm}[2]{{#1} ( #2 )}
\newcommand{\parop}{\,|\,}

\newcommand{\trs}[3]{\mbox{$\langle \hspace{-2pt} |$} #1 \; ; \; #2 \mbox{$| \hspace{-2pt} \rangle$}_{#3}}

\begin{document}

\title{On Modelling and Analysis of Dynamic Reconfiguration \\ of Dependable Real-Time Systems}


\author{\IEEEauthorblockN{Manuel Mazzara and Anirban Bhattacharyya}
\IEEEauthorblockA{Reconfiguration Interest Group\\
School of Computing Science, Newcastle University\\
Newcastle upon Tyne, UK\\
Email: \{Manuel.Mazzara, Anirban.Bhattacharyya\}@ncl.ac.uk}
}

\maketitle

\begin{abstract}
This paper motivates the need for a formalism for the modelling and analysis
of dynamic reconfiguration of dependable real-time systems.
We present requirements that the formalism must meet, and use these to evaluate
well-established formalisms and two process algebras that we have been developing, namely, Web$\pi_{\infty}$ and \textbf{CCS}$^{\lowercase{dp}}$.
A simple case study is developed to illustrate the modelling power of these two formalisms.
The paper shows how Web$\pi_{\infty}$ and \textbf{CCS}$^{\lowercase{dp}}$ represent a significant step forward in modelling adaptive and dependable real-time systems.
\end{abstract}


\section{Introduction} \label{sec:intro}

Modern dependable real-time (DRT) systems are required to have greater flexibility, availability
and dependability than their predecessors. One way of achieving this is through the use
of dynamic reconfiguration techniques.
A system can be implemented as a collection of configurations,
where each configuration is a network of communicating components.
Different configurations can be optimized for different services or operating conditions.
Therefore, facilities for replacing one configuration by another at runtime
can increase the flexibility of the system.
Runtime facilities for creating configurations that provide new services,
and for removing configurations providing services that are no longer required,
can reduce the downtime for maintenance; thereby increasing the availability of the system.
Replacing an executing configuration when it exhibits erroneous behaviour
with a valid configuration increases the system's reliability; and thereby increases its dependability.
However, dependability is a combination of reliability and predictability,
and ensuring predictability during dynamic reconfiguration is difficult.
The main reason for the difficulty is the interaction between the application services
that the system is required to provide to its environment
and the reconfiguration services that the system must also perform in order to support its
flexibility, availability and reliability. The interaction between the two kinds of service
implies that neither can be analyzed in isolation.

Existing research on DRT systems has largely concentrated on system design and programming languages
(see \cite{Kopetz97} and \cite{Burns01}) rather than on formalisms and their methods.
However, formal methods are important because they are useful in providing the strong guarantee of system correctness
required for such systems. Furthermore, the formal research on DRT systems has focused on scheduling
(see \cite{TinBurWel92}, \cite{AudBurDavSchWel96} and \cite{Mon04})
rather than on computational models.
As for the existing research on dynamic reconfiguration, it has either assumed mode changes to be instantaneous
or has implicitly assumed the controlled environment can wait whilst the control system is reconfigured
\cite{KraMag90}. Both assumptions are unrealistic for DRT systems.
For example, it is impossible to perform an instantaneous mode change
in a distributed control system because a distributed system has no global state;
and suspending or aborting application services during reconfiguration in an unstable `fly-by-wire' aircraft
would cause the aircraft to become unstable, and possibly suffer catastrophic failure.
Thus, there is very little research on computational models with overlapping modes --
the most appropriate form of dynamic reconfiguration for DRT systems.
Therefore, the purpose of our research is to develop a computational formalism for DRT systems,
in which interactions between application and runtime system activities can be modelled; 
and the model can be used to verify safety and liveness requirements of the system.
DRT systems are typically used to maintain the stability of unstable environments and to keep them
under control. Therefore, they are characterised by time-critical concurrently executing activities
with hard deadlines, small synchronisation tolerances between events, and tight resource constraints.
Hence, our formalism must be able to express both reconfiguration and real-time features of DRT systems,
and must be able to verify both their functional and timeliness properties.

\subsection*{Paper's Contribution and Structure}
 
This paper makes three contributions. First, it identifies requirements on a formalism for DRT systems that have overlapping modes. 
Second, it evaluates well-established formalisms against these requirements. 
Third, it briefly describes two novel process algebras that we believe progress the state of the art in this field.
The rest of the paper is organized as follows: section~\ref{sec:reqts} describes a simple framework
that illustrates the scope of the reconfiguration we are considering,
and then identifies the structural, modelling and analysis requirements that a formalism
targeted on the dynamic reconfiguration of DRT systems must meet.
Section~\ref{sec:forms} uses these requirements to evaluate well-established formalisms
and identifies their strengths and weaknesses. Sections~\ref{sec:webpi}~and~\ref{sec:ccsdp}
describe two formalisms we have been developing, and evaluates them against the requirements. 
Finally, Section~\ref{sec:example} describes a simple case study to illustrate the modelling power of these two formalisms.

\section{Requirements on a Formalism for Dynamic Reconfiguration} \label{sec:reqts}

\begin{figure}
  \centering
  \includegraphics[scale=0.25]{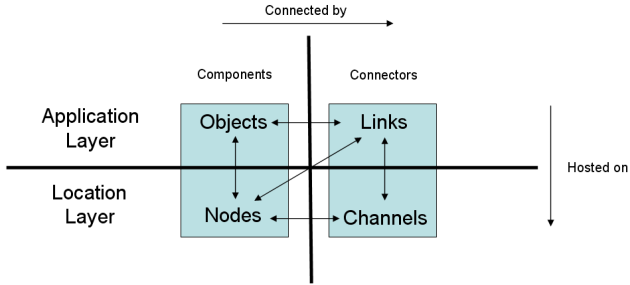}
  \caption{Framework for Dynamic Reconfiguration}
  \label{fig:conf}
\end{figure}

Reconfiguration of a DRT system typically involves changing one configuration
-- a network of communicating and concurrently executing components of the system -- into another.
Software components can also migrate between networked computers.
The computers and their network constitute the hardware platform of the system, which does not change.
We can represent the system using a simple framework (see Figure~\ref{fig:conf}).
The \textit{application layer} consists of those components of the system and their communication connectors
(i.e. \textit{objects} and \textit{links}) that are the focus of reconfiguration.
The \textit{location layer} (which does not change)
is used to represent those components and connectors (i.e. \textit{nodes} and \textit{channels})
that are necessary in order to describe the migration of objects and links.
Thus, reconfiguration can be represented as the creation and deletion of objects and links,
and as changes in the mappings between objects and links, objects and nodes, links and channels,
and links and nodes.

It is important to notice that a method for reconfiguration is not included in the framework
or in the requirements. This is because we believe that determining a method is the responsibility of the system designer, rather than the formalist.
As an analogy, the differential and integral calculus does not contain a method for designing car engines; 
but the calculus is still useful in this respect. These aspects have been properly discussed in \cite{mazzara:ewdc09}.

In order to be practicable, it must be possible to support the formalism with tools
for both modelling and automated analysis.

The requirements we have identified fall into three categories: structure, modelling and analysis. The completeness of these categories can only be determined with respect to a full case study, which is beyond the scope of this paper (in Section~\ref{sec:example} only a small example will be presented).

\subsection*{Structural Requirements}

It must be possible to model components in a recursive and compositional way.
To do this, models must be organizable in terms of units,
with each component of a system expressible as a composition of one or more units,
and each unit corresponding to only one component. This way,
it should be possible to express independent reconfiguration of units,
in the same way that components can be independently reconfigured.

\subsection*{Modelling Requirements}

The formalism must be able to express the following:

\subsubsection{Reconfiguration of Components}

it must be possible to express the creation, deletion and replacement of objects,
and also the migration of objects between nodes (see Figure~\ref{fig:conf}).

\subsubsection{Reconfiguration of Connectors}

it must be possible to express the creation and deletion of links between objects.

\subsubsection{Application Behaviour}

it must be possible to express the functionality of an application in terms of basic activities
(such I/O, data transformation and data manipulation) and their composition using sequencing,
parallelism, alternatives, iteration and recursion.

\subsubsection{Interference with Runtime Support}

it must be possible to express interference between the application
and reconfiguration activities of the system.
The interference can be functional or temporal, and can occur in both directions.
For example, functional interference can occur because the replacement of an object
can change the output of the application;
and conversely, the execution of an object that is to be replaced can change its state,
and the reconfiguration activity must take this state change into account when replacing the object.
Temporal interference is the effect on timing properties (such as execution time)
of the concurrent execution of application and reconfiguration activities;
and it can occur in the absence of functional interference.

\subsubsection{Real-Time Information}

it must be possible to model concurrent activities in terms of their ordering, duration, communication,
interrupts and memory usage for schedulability analysis. Clocks must also be modelled.

\subsubsection{Real-Time Restrictions}

it must be possible to express restrictions arising from requirements or resource limitations,
such as deadlines, computation times and bounds on synchronization.
Since memory is limited, it must be possible to express restrictions on memory usage.

\subsubsection{Fault Tolerant Behaviour}

it must be possible to express the failure modes of a system and their failure rates;
and the way in which errors will be recovered.

\subsection*{Analysis Requirements}

The main purpose of modelling is analysis. Therefore, we identified a number of analyses
that must be supported by the formalism: the formalism must be consistent.
That is, it must not be possible to prove both a statement and its negation;
otherwise, the result of an analysis can be logically invalid.
Critical properties of the system modelled using the formalism must be decidable.
That is, the function evaluating these properties must be Turing computable \cite{Turing36}.
If these two requirements are not met, it will not be possible to provide automated tool support
and the formalism will not be used in practice.

Termination of critical activities must be decidable (see `Analysis Requirements' in Section ~\ref{sec:webpi} for a discussion).
That is, it must be possible to decide whether or not any activity will stop its execution.
Deadlock detection must also be decidable.
That is, it must be possible to decide whether or not two or more processes can wait indefinitely for each other
in a cycle \cite{Zobel83}.
It must also be possible to decide whether or not activities can meet their deadlines
with respect to a scheduling discipline.
It must be possible to check whether or not data transformation and date manipulation are type correct.
It must be possible to perform reliability analysis using the failure modes and failure rates given in the model.

\section{Evaluation of Formalisms} \label{sec:forms}
The requirements described above can be used to evaluate existing formalisms for their suitability for the modelling and analysis of dynamic reconfiguration of DRT systems. Lack of space prevents a detailed and comprehensive review. Therefore, we briefly review a selection of well-established model-based formalisms, process algebras, Petri nets and other formalisms, and identify their significant strengths and weaknesses. Notice that the parts dedicated to ``Analysis Requirements''  are more focused on the existing tool support for the specific formalism, since some of the properties described above are not always applicable (e.g. deadlock for sequential formalisms) or are obviously satisfied.

\subsection{Model-based formalisms}
There is a long tradition of methods in this category. The most famous are probably VDM \cite{BjornerVDM78}, Z \cite{AbrialZ-book}, B-method \cite{AbrialB-book} and, lately, Event-B \cite{AbrialEvent-B}. The mathematics underlying these formalisms are set theory and first order logic. The approach consists of modeling the system's state in terms of sets and functions, and modelling state transformation using operations (or events in the case of Event-B). Predicates are used to express invariant conditions on the state. In Z, the emphasis is on formal specification, whilst the B-method emphasizes the ``method'' itself. Both B and Event-B focus on the application of stepwise refinement (reification in VDM). That is, the verifiable transformation of an high-level formal specification into an executable program.

\subsection*{Structural Requirements}
The VDM Specification Language (VDM-SL) and its extended form (VDM++) deal with structure in different ways: VDM-SL uses modules, whilst VDM++ (being object-oriented) uses classes with multiple inheritance. Thus, both can express structural information. Classical B and Event-B specifications are instead organized in machines that also impose a basic structure. 

\subsection*{Modelling Requirements}
Reconfiguration is not natively supported by these formalisms, but it can be encoded with difficulty (as in Turing Machines). None of these formalism has been designed for reconfiguration. Also, real-time information and restrictions are not natively supported. Temporal extensions exist (e.g. temporal logic for Z), but they do not enable all the real-time requirements to be met. Fault tolerance is not natively supported, but many contributions exist in this field. The Deploy project (www.deploy-project.eu) is addressing these issues for Event-B.

\subsection*{Analysis Requirements}
Model-based formalisms are mature, and they tend to have extensive tool support. For example, Overture for VDM (www.overturetool.org) and Rodin for Event-B (www.event-b.org). They have facilities for type checking, verifying partial correctness of a design, and checking termination. However, they are more targeted on sequential systems, and (therefore) properties characteristic of concurrent systems, such as deadlock freedom, are not directly addressed. This does not mean that they cannot be used to model concurrent systems at all. In particular, Event-B can represent interactive systems. However, events are atomic and are associated with an interleaving semantics without interference.

\subsection{Process Algebras}
Model-based formalisms are mainly concerned with functional properties and sequential behavior. In contrast, process algebras are concerned with interaction between concurrent processes. Among the original methods in this field, we can mention CSP \cite{CSP} and CCS \cite{MilnerCCS}. Mobile process algebras (e.g. Milner's $\pi$-calculus \cite{Milner99}) represent a further development by addressing mobility.

\subsection*{Structural Requirements}	
The common structural unit of all process algebras is a communicating concurrent process. Process algebras supporting a basic form of structuring do exist, although of a different nature if evaluated with respect to the requirements discussed in this paper. For example, the ambient calculus \cite{CardelliMobile} includes a notion of locations and mobility. Web$\pi_{\infty}$ is described later in this paper with the structure it imposes on the $\pi$-calculus.

\subsection*{Modelling Requirements}	
Mobile process algebras like the $\pi$-calculus are interesting because of their treatment of component bindings as first class objects, which enables link reconfiguration to be expressed simply. Although proper component reconfiguration is absent, the reconfiguration mechanism on which the $\pi$-calculus is built already represents a seminal form of what is described in this paper. Extensions to support real-time and fault tolerance are an active area of research.
	
\subsection*{Analysis Requirements}
The weakness of this category of language is tool support. Although different bisimulations have been defined and tailored to specific needs, tool support is still limited. It is worth mentioning TyPiCal, a type-based static analyzer for the $\pi$-calculus \cite{TyPiCal}. TyPiCal is able to provide four different kinds of program analyses and transformations: lock-freedom analysis (certain communications or synchronizations will eventually succeed), deadlock-freedom analysis, useless-code elimination (it removes sub-processes that do not affect the observable behavior of the process), and information flow analysis. The type system is extended in such a way that channel types carry information on how channels are used. This allows a type inferencer to obtain information about the behavior of a process. As a drawback, the expressive power of the type system is limited. 
	
\subsection{Petri Nets}	
Petri nets \cite{Petri} are a graph-based formalism to represent concurrency. They are a mathematical formalism, but they also come with an appealing graphical notation in the style of UML activity diagrams \cite{FowlerUML}. To the best of our knowledge, Petri nets were the first formalism for describing concurrency. A formal account of Petri nets in the form of a survey can be found in \cite{Manson1988}. 

\subsection*{Structural Requirements}
Structural information like modules is not natively expressible in Petri nets. Although it is a suitable formalism to express parallel and distributed systems, for a long time it did not fully support compositionality, and this deficiency prevented its wide use in large real-world applications. The recent `hype' on formalisms for verification of Web Services composition lead to some work done in this field in opposition to the process algebra approach. Indeed a big debate arose in the recent years to this regard and it is well explained in \cite{vanderAalst04} and in the conclusions of \cite{piBPEL}. Some work for enhancing Petri nets compositionality in other contexts has been done (a survey in \cite{Anisimov96}) and also work on modularization and Petri nets do exist \cite{PetriMod} but we are not aware of applications. 

\subsection*{Modelling Requirements}	
Petri nets do not offer a native way for addressing dynamic reconfiguration, but extensions to the formalism have been presented to allow for an easy formalization of this feature. For example, reconfigurable Petri nets \cite{Llorens04} are a subclass of net rewriting systems with the goal of enhancing the expressiveness of the basic model. Other approaches to model dynamic reconfiguration have been tried and shown through case studies \cite{Lemmin00}. The original version of Petri nets is not Turing complete but extension have been provided later to add expressiveness.
	
\subsection*{Analysis Requirements}		

Tool support for Petri nets benefited from decades of research on the topic (see www.informatik.uni-hamburg.de/TGI). Various kinds of Petri net are supported by tools, and many of these tools offer a practical and appealing graphical editor not offered by other formalisms. Animation and model checking are other useful features offered by some of these tools. Overall, it is probably not an exaggeration to say that, in comparison to process algebras, tool support for Petri nets have received much more attention from the scientific community. 

\subsection{Other Formalisms}		
We are aware of others formalisms that deserve attention. Here we will briefly mention some of those with their main features.

The Chemical Abstract Machine (CHAM) \cite{Gerard92} exploits the chemical metaphor. That is, it follows an approach based on viewing software systems as chemicals whose reactions are rigorously controlled by specific rules. The original idea was to bridge the gap between Petri nets, which can be considered as abstract machines but lack expressiveness (in the basic version) and process algebras, which are more expressive but are intended as specification formalisms for distributed systems (rather than as abstract machines). It is also worth mentioning how the gap between calculi for concurrent processes and languages for programming distributed and mobile systems is not really bridged by CCS or $\pi$-calculus, since their interactions are based on rendezvous, i.e. atomic and non-local, which is hard to implement fully in a distributed setting. Here is where CHAM finds its niche and it is very interesting for its ability to represent a system as a syntactic description of the static components, the molecules, and of a set of reaction rules describing how the system evolves dynamically. This already shows how a system is structured in the formalism and how some form of reconfiguration can be expressed. Furthermore, formal reasoning, for example about deadlocks, can be performed.   

Another attempt of combining pros and cons of different formalisms and bridging the gap between them is CSP$\parallel$B \cite{SchneiderT05}. The authors recognize that a system can be projected into two different dimensions, the dynamic view and the state view, and these projections have to be consistent. Thus, they combine B with a \textit{controller language} able to \textit{drive} a B machine and this controller language is (a subset of) CSP. In this way B is able to express requirements on the state of a system while CSP expresses the interactive behaviour. This approach should permit the exploitation of existing tool support for both CSP and B. CSP$\parallel$B is interesting for the purpose of this paper since it combines the structure of B and it has the potential of exploiting mobility like the $\pi$-calculus, although not in the basic version \cite{Schneider07}.

Notable tool support in this category is provided by UPPAAL -- an integrated tool environment for modelling, validation and verification of real-time systems, modelled as networks of timed automata extended with data types (www.uppaal.com).


	
\section{$Web\pi_{\infty}$} \label{sec:webpi}

In \cite{MazzaraPhDThesis} and \cite{MazzaraL06}, a unifying theory has been developed with the pragmatic intention of using it for encoding orchestration languages behavior (WS-BPEL in particular) and verifying process equivalence. Certainly, the reconfigurability needs in that scenario are limited compared to the general case: processes can be rolled-back or compensated, fault handlers activated, but they cannot be dynamically deleted or created. Both in WS-BPEL and in the developed theory Web$\pi_{\infty}$ there is always the need of statically defining a "syntactical proximity" between the process responsible for the normal behavior and the one responsible for the `abnormal' one. This means that the two processes have to be statically bound. More complex behavior can be certainly encoded and the problem circumvented but we still have noticed that this practice would represent a sort of unpleasant `hacking' that it is still far from what we want to offer to the final user. Despite this, we still recognize one step forward with respect to previous formalisms for what concern inborn reconfigurable behavior. First, being  Web$\pi_{\infty}$ based on the $\pi$-calculus, it allows the same sort of flexibility of its ancestors, i.e. link passing. But it does more offering a reconfigurable behavior as a first class citizen. In fact, when compared to the ideal formalism --- the cornucopia able to elegantly satisfy all the presented requirements --- Web$\pi_{\infty}$ appears still primitive but it contains, in a seminal form, many of those requirements.

Web$\pi_{\infty}$ is a conservative extension of the $\pi$-calculus where the workunit operator $\trs{P}{Q}{x}$ has been added. Here the normal behavior is expressed by the process $P$ and the abnormal one by $Q$, while $x$ is the ``trigger'' that allows to switch from one to the other. This ``trigger'' is able to activate $Q$ during the execution of $P$ if another parallel process $\opm{x}{}$ (output) requires it. The evolution of the parallel composition of an output and a workunit (according to the formal reduction semantics) is:  $\opm{x}{} \parop \trs{P}{Q}{x} \rightarrow \trs{Q}{\pinull}{}$ that, for technical reasons here omitted, will then behave like $Q$.

We believe this formalism shows some elegance and it is able to integrate structural, modelling and analysis requirements.





\subsection*{Structural Requirements}
Web$\pi_{\infty}$ expresses information on the structure being the system organized in different workunits. The basic structural unit is a process, and workunits are used to perform reconfiguration representing what is being changed and the change. They can be recursive (nested) and compositional. Each of the workunit could, ideally, reside in a different host and could be compiled and linked separately. Although  it is a very basic mechanism for structural information it represents an improvement in comparison to the $\pi$-calculus.

\subsection*{Modelling Requirements}	

\subsubsection{Reconfiguration of Components}
reconfiguration of components is possible through workunits and handler activation. In a Web$\pi_{\infty}$ process behaviour and reconfiguration are represented by $P$ (application) and the interaction between $\opm{x}{}$ and the workunit, which produces the execution of the process $Q$ ``replacing'' $P$ (reconfiguration).

\subsubsection{Reconfiguration of Links}
reconfiguration of links is inherited by the $\pi$-calculus and its notion of mobility (which is a basic form of dynamic reconfiguration). The $\pi$-calculus looks interesting because of its treatment of component bindings as first class objects, which enables dynamic reconfiguration to be expressed simply.

\subsubsection{Application Behaviour}
the functionality of an application in terms of basic activities are abstracted over in Web$\pi_{\infty}$. This feature is inherited from the $\pi$-calculus which has the purpose of representing process synchronization, communication and link mobility. Activities like sequence, parallel, alternative, iteration, etc. are expressible through encoding. This exercise has been done in \cite {piBPEL} to encode WS-BPEL. This shows how  Web$\pi_{\infty}$ is expressive enough to be used to describe the application behavior.

\subsubsection{Interference with Runtime Support}
interference between the application and reconfiguration activities of the system is expressible through message passing and rendez-vous, as shown in the example above. The interference can only be functional, not temporal.

\subsubsection{Real-Time Information}
it is possible to model concurrent activities in terms of their ordering and communication, but duration is not expressible (at least in the untimed version). Memory usage is not expressible, whilst interrupts are natively supported. Clocks cannot be modelled.

\subsubsection{Real-Time Restrictions}
Web$\pi_{\infty}$ is not able to express restrictions arising from time and space limitation limitations. However, a timed version of the language has been presented (but not coping with real time).

\subsubsection{Fault Tolerant Behaviour}
fault tolerance is a point in  Web$\pi_{\infty}$ when describing, for example, the WS-BPEL recovery framework \cite{MazzaraBPEL2009} and it has been one of the main reason for its development. Failure rates are not expressible.

\subsection*{Analysis Requirements}	
	
Web$\pi_{\infty}$ is a conservative extension of the $\pi$-calculus and can be encoded using it. The reason for developing a new formalism was not the nature of its expressiveness, but its pragmatics. As a consequence, consistency is inherited from the $\pi$-calculus. Termination, in the general case, is not decidable for Turing-complete formalisms, including the $\pi$-calculus. In practical cases, and with the necessary restrictions (e.g. by typing and syntax), termination can be ensured. An adequate discussion on this topic is in \cite{Sangiorgi2006}. The same holds for Web$\pi_{\infty}$. Decidability of termination is a theoretically well-known limit; when designing a formalism it is always a matter of practicality finding a compromise between the expressiveness of the languages and the properties that can be decided.

Looking at the other requirements, in Web$\pi_{\infty}$ it is not possible to decide whether or not the activities can meet their deadlines with respect to a scheduling discipline. It is not possible to check whether or not data transformation and manipulation is type correct. It is not possible to perform reliability analysis using failure modes and rates. Web$\pi_{\infty}$ comes with its tailored definition of bisimulation, i.e. a mathematical tool able to determine if two processes exhibit the same externally visible behavior. The proposed bisimulation is decidable for non-recursive processes and some properties are proved, as an example, in  \cite{MazzaraPhDThesis}.

\section{$CCS^{\lowercase{dp}}$} \label{sec:ccsdp}

$CCS^{dp}$ is the first version of a formalism that is being developed specifically for
the modelling and analysis of interactions between application and reconfiguration activities
in DRT systems \cite{BhaFit08}. It is based on $CCS$ \cite{MilnerCCS}, extended with a single construct --
the fraction process $\frac{P'}{P}$ -- in order to reconfigure processes.

A fraction process $\frac{P'}{P}$ is used to replace and delete processes.
On creation, the fraction $\frac{P'}{P}$ identifies any instance of a process
matching the denominator process $P$ with which it is composed in parallel,
and replaces that process immediately and atomically with the numerator process $P'$.
The matching can be either syntactic equality or
a special kind of behavioural equivalence ($\sim_{of}$) we have defined.
The reduction semantics is given by $P | \frac{P'}{P} \longrightarrow P' \;$
and $\; Q | \frac{P'}{P} \longrightarrow P'$ where $P \!\!\sim_{of}\!\! Q$.
If no matching process instance exists,
the fraction continues to exist until such a process is created
(or the fraction is itself deleted or replaced).
If there is more than one matching process instance,
a non-deterministic choice is made as to which process is replaced.
Similarly, if more than one fraction can replace a process instance,
a non-deterministic choice is made as to which fraction replaces the process.
Deletion of a process P is achieved by parallel composition with \mbox{$\frac{0}{P}$}.
If $P$ progresses to $R$, then \mbox{$\frac{P'}{P}$} will not replace $R$ by $P'$
(unless $R$ matches $P$). Notice that a fraction process has no intrinsic behaviour;
it performs a transition only when composed with a process that matches its denominator.

The key strengths of $CCS^{dp}$ regarding dynamic reconfiguration of DRT systems are:
its ability to separate and compose models of application and reconfiguration activities,
resulting in modular and terse models; expressing application and reconfiguration actions
in the same form, so that their interleaving can be easily represented;
and its simplicity in modelling process reconfiguration.
The key weaknesses of $CCS^{dp}$ are: its lack of facilities for naming processes,
and for modelling real-time and fault tolerance properties of a system.

\subsection*{Structural Requirements}

The basic structural unit is a process.
The parallel composition operator ($|$) of $CCS$ enables a process to be composed from parallel processes,
and a process to be decomposed into parallel processes.
The restriction operator ($\nu$) facilitates process composition by scoping port names.
These facilities are inherited by $CCS^{dp}$.
The semantics of fraction processes enables application and reconfiguration activities
to be modelled separately, and then composed using $|$ to enable reconfiguration to take place.
Thus, fractions support modular structuring of a model.
However, since there is no naming scheme for processes,
identical instances of a process cannot be selectively reconfigured.

\subsection*{Modelling Requirements}

\textit{1) Reconfiguration of Components:}
replacement and deletion of components are expressed as reconfiguration transitions
involving a fraction process. Component creation is expressed as process spawning (as in $CCS$).
The granularity of process reconfiguration is a concurrent process,
and any parallel composition of processes can be reconfigured. Process reconfiguration is atomic.
Process migration cannot be modelled, since the location of a process is not represented.

\textit{2) Reconfiguration of Links:}
this can be expressed using process reconfiguration; but it is clumsy.

\textit{3) Application Behaviour:}
facilities for expressing this are inherited from $CCS$.
A process can perform I/O actions (without value-passing),
internal action, sequential action, iterative action,
and make a deterministic or non-deterministic choice between alternative actions.
Concurrent processes execute with interleaved semantics and communicate synchronously.

\textit{4) Interference with Runtime Support:}
this is expressed in terms of interleavings between the application transitions and the reconfiguration transitions
of a process expression, and the resulting process expression shows the outcome of the interference.
Temporal interference cannot be described, since $CCS^{dp}$ has no time model.

\textit{5) Real-Time Information:}
this is inherited from $CCS$, and is limited.
Concurrent processes and the order of actions can be expressed, but not the duration of an action
(which excludes schedulability analysis).
The synchronous communication between processes is also unsuitable for DRT systems.
Memory usage, interrupts and clocks are not modelled.

\textit{6) Real-Time Restrictions:}
these cannot be expressed in $CCS^{dp}$.

\textit{7) Fault Tolerant Behaviour:}
there are no special facilities in $CCS^{dp}$ to express this.

\subsection*{Analysis Requirements}

Work on the analytical aspects of $CCS^{dp}$ is in progress:
proof of consistency is significant
because of negative premises in some of the semantic rules of $CCS^{dp}$.
Proof of decidability of the $\sim_{of}$ bisimulation is significant for matching.
It is important to identify a practically useful set of processes for which we can prove
decidability, termination and deadlock freedom. However, schedulability analysis,
type checking and reliability analysis are beyond the scope of $CCS^{dp}$.

\section{Case Study} \label{sec:example}

In this section, we illustrate the modelling power of Web$\pi_{\infty}$ and $CCS^{dp}$ 
by means of a simple case study: a `stripped-down' sensor array (see Figure 2).

The sensor array consists of a number of identical hardware sensors,
each of which is handled by a separate software process; and a reconfiguration manager.
To maximize the longevity of the array, only one sensor is active at a time;
the other sensors are either dormant or `burned-out'.
The array operates by the software process of the active sensor sending its reading to the reconfiguration manager,
which processes the reading.
If the sensor starts to `burn-out', it intermittently outputs an error signal that causes the reconfiguration manager to reconfigure the array
by deleting the faulty sensor's software process, and creating a new one to handle a newly activated hardware sensor.
All the software processes are non-terminating.

\begin{figure}[hb]
\centering
\includegraphics[width=0.48\textwidth]{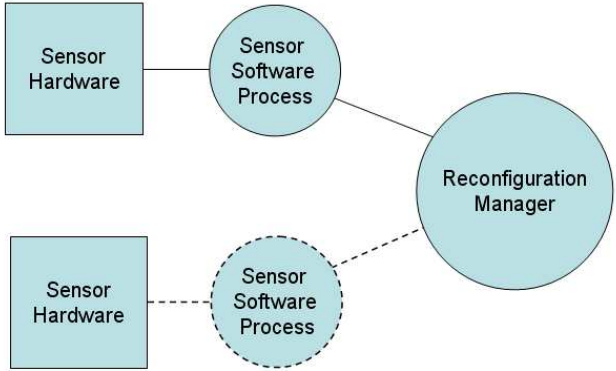}
\caption{Simplified Sensor Array} \label{FigSensorArray}
\end{figure}

\subsection*{Web$\pi_{\infty}$ Model}

Let $S$ be a sensor software process. $S$ behaves nondeterministically -- either correctly or erroneously. 
In the first case, $S$ will end up executing itself (after `garbage collection' of pending messages); 
whilst in the second case, $S$ will send a message $\opm{e'}$ (again after ``garbage collection'').
We define $S$ in Web$\pi_{\infty}$ as follows:
$S \triangleq \ipm{v}{}.\ipm{e}{}.S + \ipm{e}{}.\ipm{v}{}.\opm{e'}$ 
Let $R$ be the reconfiguration manager. $R$ consists of a workunit with body $S$. 
In the erroneous case, the handler of the workunit in $R$ will be activated and restore $S$ itself, which in turn, can still behave correctly or erroneously. 
We define $R$ in Web$\pi_{\infty}$ as follows: 
$R \triangleq \trs{S}{S}{e'} \parop \opm{v} \parop  \opm{e}$

The reductions in both the normal and erroneous cases are:
\begin{flushleft}
\textit{Normal case}:
$\trs{\ipm{v}{}.\ipm{e}{}.S + \ipm{e}{}.\ipm{v}{}.\opm{e'}}{S}{e'} \parop \opm{v} \parop \opm{e}$
$\rightarrow \trs{\ipm{e}{}.S}{S}{e'} \parop \opm{e} \rightarrow \trs{S}{S}{e'}$
\end{flushleft}
\begin{flushleft}
\noindent \textit{Erroneous case}:
$\trs{\ipm{v}{}.\ipm{e}{}.S + \ipm{e}{}.\ipm{v}{}.\opm{e'}}{S}{e'} \parop \opm{v} \parop \opm{e}$ 
$\rightarrow \trs{\ipm{v}{}.\opm{e'}}{S}{e'} \parop \opm{v} \rightarrow \trs{\opm{e'}{}}{S}{e'} \rightarrow \trs{S}{\pinull}{}$
\end{flushleft}

\subsection*{$CCS^{dp}$ Model}

Let $S$ be a sensor software process. $S$ behaves either correctly (performing $\bar{v}$)
or incorrectly (performing $\bar{e}$). We can define $S$ as follows: $S \triangleq \bar{v}.S + \bar{e}.S$ \\
Let $R$ be the reconfiguration process. $R$ either lets $S$ continue (when $S$ behaves correctly),
or replaces it with a different instance of $S$ (when $S$ behaves incorrectly).
$R$ is defined as follows: $R \triangleq v.R + e.(\frac{S|R}{S})$

The reductions in both the normal and erroneous cases are:
\begin{flushleft}
\textit{Normal case}:
\(
S | R \longrightarrow
S | R
\)
\end{flushleft}
\begin{flushleft}
\noindent \textit{Erroneous case}:
\(
S | R \longrightarrow
S | \frac{S|R}{S} \longrightarrow S | R
\)
\end{flushleft}

\section{Conclusions and Future Work} \label{sec:concl}

%

In this paper, we have focused on reconfiguration with interference between
application activities and reconfiguration activities in DRT systems.
We have identified requirements on a computational formalism,
and evaluated well-established formalisms (as well as our own) against these requirements.
We have shown that none of the existing formalisms provides full native support for dynamic reconfiguration.
Some implicitly contain specific features that can be useful when treating systems that inherently show reconfigurable features; but none of them are entirely suitable for this category of problems (see Table 1).

\begin{center}
\begin{table*}[ht]
{
\hfill{}
\begin{tabular}{|l||c|c|c|}
\hline
\textbf{Formalism} & \textbf{Structural granularity} & \textbf{Support for Dynamic Reconfiguration} & \textbf{Tool support}\\
\cline{2-4}
\hline
\hline
\textbf{Event-B} & Machine & Not supported & Rodin \\
\hline
\textbf{CCS} & Concurrent process & Process creation & Edinburgh Concurrency Workbench \\
\hline
$\pi$\textbf{-calculus} & Concurrent process & Process creation; link passing & TyPiCal \\
\hline
\textbf{Petri nets} & Net & Not supported & Extensive tool support \\
\hline
\textbf{CSP}$\parallel$\textbf{B} & Machine & Process creation, deletion, replacement; link passing & Tools for CSP and B\\
\hline
\textbf{CCS}$^{dp}$ & Concurrent process & Process creation, deletion, replacement & None \\
\hline
\textbf{Web}$\pi_{\infty}$ & Concurrent process & Process creation, deletion, replacement; link passing & None \\
\hline\end{tabular}}
\hfill{}
\caption{State of the Art of Formalisms for Dynamic Reconfiguration}
\label{tab:summary}\end{table*}
\end{center}

Other formalisms could also have been evaluated,
but we decided to refer to Wermelinger's PhD thesis \cite{wermelinger99phd}, which shares our conclusion.
Wermelinger takes the premise that a single formalism can never satisfy all the requirements in every situation.
Therefore, he presents three approaches -- each one making use of a different formalism.
Each approach has its own assumptions about the system, and each has its advantages and disadvantages.

As self-criticism, we need to mention some problems with this paper that deserve further work:
first, it has been asked how we can know that the list of requirements is complete.
In fact, the completeness of a list of requirements can never be determined for a generic system, but only for a specific system.
Second, the survey of formalisms is incomplete. This was unavoidable due to space restrictions.

For future work, the members of the Reconfiguration Interest Group (RIG) at Newcastle 
are interested in exploring different aspects of dynamic reconfiguration. In order to achieve 
integration of our research, we will need a common framework.  
This framework should be formal in order to support dependability during dynamic reconfiguration: 
it should be able to model architectural configuration; express policies
that must hold for a configuration; reason about properties of the configuration -- 
for example, formally verify whether or not a policy holds
for the configuration; model the process through which a system is reconfigured; 
and verify whether or not the process satisfies the safety and
liveness requirements of the system defined over the reconfiguration interval.

\subsection*{Acknowledgments}
\noindent This work is partly funded by the EPSRC under the terms of a graduate studentship. The paper has been improved by useful conversations with Gudmund Grov, Jeremy Bryans, John Fitzgerald, Cliff Jones and Michele Mazzucco. We also want to thank members of the Reconfiguration Interest Group (in particular, Kamarul Abdul Basit, Carl Gamble and Richard Payne),
the Dependability Group (at Newcastle University) and the EU FP7 DEPLOY Project (Industrial deployment of system engineering methods providing high dependability and productivity).


\bibliographystyle{unsrt}
\bibliography{DEPENDrefs}


\appendix
\section*{Summary of Requirements}

\subsection*{Model Structuring Requirements}
Unit encapsulation, unit compositionality, unit naming.

\subsection*{System Modelling Requirements}
\begin{itemize}
	\item \textit{Reconfiguration of Components}: component creation, component deletion, component replacement, component migration.
	\item \textit{Reconfiguration of Links}: link creation, link deletion.
	\item \textit{Application Behaviour}: input, output, calculation, data model, state update, control structures (sequence, parallel, alternative, iterations, recursion).
	\item \textit{Interference with Runtime Support}: functional interaction, temporal interaction.
	\item \textit{Real-Time Information}: time model, memory model, communication model, concurrency model, interrupt model.
	\item \textit{Real-Time Restrictions}: scheduling restrictions, synchronization restrictions, memory restrictions.
	\item \textit{Fault Tolerant Behaviour}: fault models, error recovery.
\end{itemize}

\subsection*{Analysis Requirements}
Consistency, decidability, termination, deadlock freedom, scheduling feasibility, type checking, reliability analysis.

\end{document}